\documentclass[sigconf]{acmart}

\renewcommand\footnotetextcopyrightpermission[1]{} % remove copyright footnote
\AtBeginDocument{%
  }

\usepackage{subcaption}
\usepackage{graphicx} 
\usepackage[most]{tcolorbox}
\usepackage[table,xcdraw]{xcolor} % For coloring rows and cells
\usepackage{algorithm}
\usepackage{algpseudocode}
\usepackage{amsmath}
\usepackage{graphicx} % in preamble
\begin{document}

%%
%% The "title" command has an optional parameter,
%% allowing the author to define a "short title" to be used in page headers.
\title{EAGER: Edge-Aligned LLM Defense for Robust, Efficient, and Accurate Cybersecurity Question Answering}

\author{Onat Gungor}
\authornote{Both authors contributed equally to this research.}
\affiliation{
  \institution{University of California, San Diego}
  \department{Computer Science and Engineering}
  \city{La Jolla}
  \state{CA}
  \country{USA}
}
\email{ogungor@ucsd.edu}

\author{Roshan Sood}
\authornotemark[1]
\affiliation{
  \institution{University of California, San Diego}
  \department{Computer Science and Engineering}
  \city{La Jolla}
  \state{CA}
  \country{USA}
}
\email{rosood@ucsd.edu}

\author{Jiasheng Zhou}
\affiliation{
  \institution{University of California, San Diego}
  \department{Computer Science and Engineering}
  \city{La Jolla}
  \state{CA}
  \country{USA}
}
\email{rjzhou@ucsd.edu}

\author{Tajana Rosing}
\affiliation{
  \institution{University of California, San Diego}
  \department{Computer Science and Engineering}
  \city{La Jolla}
  \state{CA}
  \country{USA}
}
\email{tajana@ucsd.edu}

%%
%% The "author" command and its associated commands are used to define
%% the authors and their affiliations.
%% Of note is the shared affiliation of the first two authors, and the
%% "authornote" and "authornotemark" commands
%% used to denote shared contribution to the research.

%%
%% By default, the full list of authors will be used in the page
%% headers. Often, this list is too long, and will overlap
%% other information printed in the page headers. This command allows
%% the author to define a more concise list
%% of authors' names for this purpose.
% \renewcommand{\shortauthors}{Trovato et al.}
\newcommand{\Design}[0]{\textsc{EAGER}}

%%
%% The abstract is a short summary of the work to be presented in the
%% article.
\begin{abstract}
Large Language Models (LLMs) are highly effective for cybersecurity question answering (QA) but are difficult to deploy on edge devices due to their size. Quantization reduces memory and compute requirements but often degrades accuracy and increases vulnerability to adversarial attacks. We present \Design{}, an edge-aligned defense framework that integrates parameter-efficient quantization with domain-specific preference alignment to jointly optimize efficiency, robustness, and accuracy. Unlike prior methods that address these aspects separately, \Design{} leverages Quantized Low-Rank Adaptation (QLoRA) for low-cost fine-tuning and Direct Preference Optimization (DPO) on a self-constructed cybersecurity preference dataset, eliminating the need for human labels. Experiments show that \Design{} reduces adversarial attack success rates by up to 7.3$\times$ and improves QA accuracy by up to 55\% over state-of-the-art defenses, while achieving the lowest response latency on a Jetson Orin, demonstrating its practical edge deployment.
\end{abstract}

%%
%% The code below is generated by the tool at http://dl.acm.org/ccs.cfm.
%% Please copy and paste the code instead of the example below.


%%
%\begin{CCSXML}
%<ccs2012>
% <concept>
%  <concept_id>00000000.0000000.0000000</concept_id>
%  <concept_desc>Do Not Use This Code, Generate the Correct Terms for Your Paper</concept_desc>
%  <concept_significance>500</concept_significance>
% </concept>
% <concept>
%  <concept_id>00000000.00000000.00000000</concept_id>
%  <concept_desc>Do Not Use This Code, Generate the Correct Terms for Your Paper</concept_desc>
%  <concept_significance>300</concept_significance>
% </concept>
% <concept>
%  <concept_id>00000000.00000000.00000000</concept_id>
%  <concept_desc>Do Not Use This Code, Generate the Correct Terms for Your Paper</concept_desc>
%  <concept_significance>100</concept_significance>
% </concept>
% <concept>
%  <concept_id>00000000.00000000.00000000</concept_id>
%  <concept_desc>Do Not Use This Code, Generate the Correct Terms for Your Paper</concept_desc>
%  <concept_significance>100</concept_significance>
% </concept>
%</ccs2012>
%\end{CCSXML}

%\ccsdesc[500]{Do Not Use This Code~Generate the Correct Terms for Your Paper}
%\ccsdesc[300]{Do Not Use This Code~Generate the Correct Terms for Your Paper}
%\ccsdesc{Do Not Use This Code~Generate the Correct Terms for Your Paper}
%\ccsdesc[100]{Do Not Use This Code~Generate the Correct Terms for Your Paper}

%%
%% Keywords. The author(s) should pick words that accurately describe
%% the work being presented. Separate the keywords with commas.
%% A "teaser" image appears between the author and affiliation
%% information and the body of the document, and typically spans the
%% page.

%\received{20 February 2007}
%\received[revised]{12 March 2009}
%\received[accepted]{5 June 2009}

%%
%% This command processes the author and affiliation and title
%% information and builds the first part of the formatted document.
\maketitle

\section{Introduction}
Cybersecurity professionals face increasingly sophisticated threats that demand accurate and timely decision-making~\cite{dekker2024threat}. In 2024, large enterprises allocated an average of \$14.6 million to Security Operations Centers (SOCs), with roughly 80\% devoted to labor~\cite{kpmg2024cybersecurity}, highlighting the need for scalable, intelligent solutions. Large Language Models (LLMs) offer a promising approach by enhancing reasoning and automation in cybersecurity workflows~\cite{tian2025exploring,ferrag2024generative}. One impactful application is cybersecurity question answering (QA), where LLMs generate context-aware, human-like responses to security queries, facilitating faster threat detection and remediation~\cite{agrawal2024cyberq,rajapaksha2024rag}. By automating QA, LLMs can help reduce the labor burden in SOCs while improving response speed and decision quality~\cite{gungor2025aqua}.

The need for rapid and context-aware decision-making is especially critical in edge environments, where timely responses and data privacy are paramount. For instance, in power grids, LLM-based QA systems could enable edge devices to interpret intrusion alerts locally, offering timely and actionable guidance. However, state-of-the-art cybersecurity QA methods~\cite{liu2023secqa, rajapaksha2024rag, tihanyi2024cybermetric} rely on GPT variants, whose substantial compute requirements hinder edge deployment. Techniques such as model optimization, edge-cloud collaboration, and hardware acceleration have been proposed~\cite{semerikov2025llm}, and quantization is particularly promising for reducing memory and compute overhead. Yet, quantization often leads to reduced accuracy and increased vulnerability to adversarial attacks~\cite{shen2024agile, zhang2024edge}.

%%%%%%%%%%%%%%%%%%%%%%%%%%%%%%%%%%%%%%%%%%%%%%%%%%%%%%%%%%%%%%%%
\begin{figure}[]
    \centering
    \includegraphics[width=0.35\textwidth]{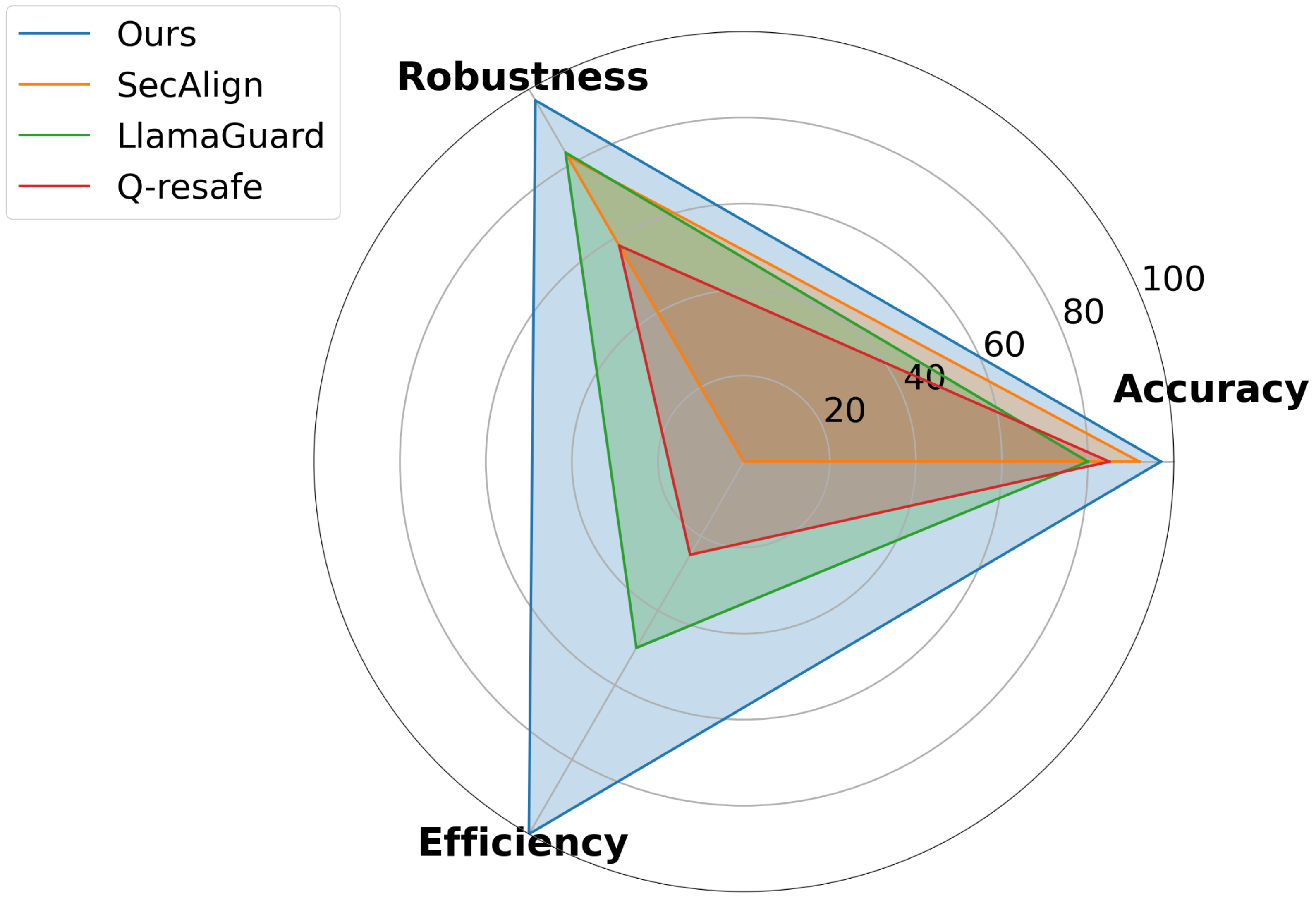}
    \caption{Comparison of existing defenses SecAlign~\cite{chen2025meta}, Q-resafe~\cite{chenassessing}, LlamaGuard~\cite{inan2023llama}, and our method (\Design{})}
    \label{fig:spider-chart}
\end{figure}
%%%%%%%%%%%%%%%%%%%%%%%%%%%%%%%%%%%%%%%%%%%%%%%%%%%%%%%%%%%%%%%%

Quantized LLMs are particularly susceptible to prompt injection attacks, where adversaries embed malicious instructions that override intended model behavior~\cite{li2024investigating, egashira2024exploiting, kumar2024fine, chandra2024adversarial, chen2024secalign}. Existing defenses~\cite{inan2023llama, chen2025meta, chenassessing} address aspects of this threat but do not simultaneously balance efficiency, robustness, and accuracy; most methods target only one or two objectives. As illustrated in Figure~\ref{fig:spider-chart}, which reports results on the CTIBench dataset~\cite{alam2024ctibench}, prior defense methods improve one or two dimensions while sacrificing the others. These trade-offs occur because efficiency, robustness, and accuracy interact: aggressive quantization can reduce robustness, while strong alignment mechanisms increase compute or memory overhead, limiting edge deployment. Importantly, these conflicts are not inherent: a carefully co-designed framework can preserve critical model representations, enhance robustness, and maintain QA accuracy simultaneously. This raises the central question: Can we design a defense that achieves high efficiency, strong robustness, and reliable QA accuracy concurrently in quantized LLMs?

%  showing their relative performance in efficiency, robustness, and accuracy. Our approach achieves balanced performance across all three dimensions

% Prior methods focus on one or two dimensions at the cost of others, while \Design{} achieves balanced performance across all three.

To address this challenge, we propose Edge-Aligned LLM Defense for Robust, Efficient, and Accurate Cybersecurity QA (\Design{}), a framework that, to our knowledge, is the first to co-design and balance efficiency, robustness, and QA accuracy for quantized LLMs. Unlike prior methods that improve one or two objectives at the expense of the others, \Design{} integrates quantization-aware fine-tuning with lightweight preference alignment in a unified framework, preserving critical model representations, strengthening robustness to prompt injection attacks, and maintaining QA accuracy simultaneously. Our key contributions are threefold:

\begin{itemize}
    \item \textbf{Co-design framework:} We integrate low-cost, quantization-aware fine-tuning (using QLoRA) with preference alignment, enabling efficient deployment with minimal overhead.
    
    \item \textbf{Domain-specific preference data:} We introduce a pipeline that self-generates cybersecurity-specific preference pairs, enabling robust alignment against prompt injection attacks via DPO without costly human-expert labeling.

    \item \textbf{Synergistic alignment:} \Design{} jointly aligns quantization and preference tuning to enhance task-specific QA accuracy, adversarial robustness, and efficiency, achieving a balance not demonstrated by prior work (Figure~\ref{fig:spider-chart}).

\end{itemize}

Experimental results show that \Design{} reduces adversarial attack success rates by up to 7.3$\times$ (average 4.9$\times$), improves QA utility by up to 55\%, and provides the lowest response latency on a Jetson Orin compared to state-of-the-art defenses~\cite{chen2024secalign}. By balancing all three aspects simultaneously, \Design{} offers a practical, edge-aligned solution for robust and efficient cybersecurity QA.

\section{Background and Related Work}
\textbf{LLM-based Cybersecurity Question Answering.} Cybersecurity QA tasks evaluate a model’s ability to provide accurate, contextually relevant answers. LLMs leverage broad technical knowledge to reason across diverse cybersecurity scenarios. Most state-of-the-art LLM-based cybersecurity QA solutions~\cite{liu2024cyberbench,tihanyi2024cybermetric,liu2023secqa,zhao2024ontology,alam2024ctibench} primarily benchmark different models and improve predictive performance. CyberLLM-Instruct~\cite{eizemity2025cyberllminstruct} and AQUA-LLM~\cite{gungor2025aqua} are two notable studies that examine the safety of LLM-based cybersecurity QA systems and characterize different types of attacks, yet neither proposes defense mechanisms. Furthermore, these approaches mostly rely on large-scale models, such as GPT variants, and do not consider optimization techniques like quantization, which are critical for deployment on resource-constrained edge devices.

\textbf{Prompt Injection Attacks on LLMs.} Prompt injection, identified by the 2025 OWASP Top Ten for LLMs as the most critical cybersecurity threat~\cite{owasp2025llmtop10}, exploits a model’s instruction-following behavior to manipulate outputs at inference time. Such attacks are classified as direct—where malicious input is explicitly provided by the user—or indirect, originating from external sources such as web pages~\cite{chen2024secalign}. We focus on direct prompt injections, which pose the most immediate threat. This vulnerability is particularly critical in cybersecurity QA systems, where carefully crafted inputs can mislead the model into producing harmful or incorrect guidance.

\textbf{LLM Alignment.} Reinforcement Learning from Human Feedback (RLHF) guides LLMs to generate outputs aligned with human judgments, improving adherence to desired behaviors~\cite{ouyang2022training}. Existing approaches train on human-labeled comparisons, rank outputs by quality, or use reward models to guide generation. Direct Preference Optimization (DPO) bypasses reward modeling by directly matching model behavior to human choices~\cite{rafailov2023direct}, reducing training overhead while producing outputs aligned with human preferences, making it an efficient and practical RLHF solution.

\textbf{Prompt Injection Defenses.} Defenses against prompt injection are grouped into three categories. Fine-tuning–based methods enhance model robustness by adjusting the model’s parameters~\cite{piet2024jatmo}, detection-based approaches filter malicious instructions before they are processed by the LLM~\cite{inan2023llama}, and prompting-based strategies steer the model’s outputs using carefully crafted prompts~\cite{chen2024defense}. Empirical studies suggest that fine-tuning approaches generally achieve the highest robustness~\cite{chen2025struq}, with representative defenses including Jatmo~\cite{piet2024jatmo}, StruQ~\cite{chen2025struq}, and ISE~\cite{wu2024instructional}. However, SecAlign~\cite{chen2024secalign} showed that these methods do not explicitly discourage undesirable outputs, limiting their overall effectiveness. To address this, SecAlign frames prompt injection defense as a preference optimization problem. Although SecAlign demonstrates strong robustness against prompt injections, it is not designed for resource-constrained edge deployment and depends on large human-labeled datasets that are not tailored to the cybersecurity domain. These limitations highlight the need for efficient and domain-specific defenses.

\textbf{Safety of Quantized LLMs.} Quantization has become a key technique for deploying LLMs on resource-constrained edge devices~\cite{xiao2023smoothquant}. By converting high-precision weights (e.g., 16-bit) into lower-precision formats such as 4-bit, quantization reduces memory and computational requirements while maintaining comparable model performance. However, recent studies show that quantization can compromise LLM safety, increasing the risk of harmful or unintended behaviors under adversarial inputs~\cite{li2024investigating, kumar2024fine, hong2024decoding, egashira2024exploiting, chandra2024adversarial}. Q-resafe~\cite{chenassessing} was recently proposed to restore the safety behavior of quantized LLMs, re-aligning safety-critical weights with their pre-quantization counterparts. While promising, Q-resafe relies on preference data from external models rather than the target model, requires costly weight updates via stochastic gradient descent, lacks comparisons with state-of-the-art defenses, and provides no evidence of efficiency for edge deployment. In contrast, our framework uses domain-specific cybersecurity preference data to achieve more effective alignment while maintaining edge efficiency.

\section{\Design{} Framework}
\label{framework}
%%%%%%%%%%%%%%%%%%%%%%%%%%%%%%%%%%%%%%%%%%%%%%%%%%%%%%%%%%%%%%%%%%%%%%
\begin{figure}[]
    \centering
    \includegraphics[width=0.48\textwidth]{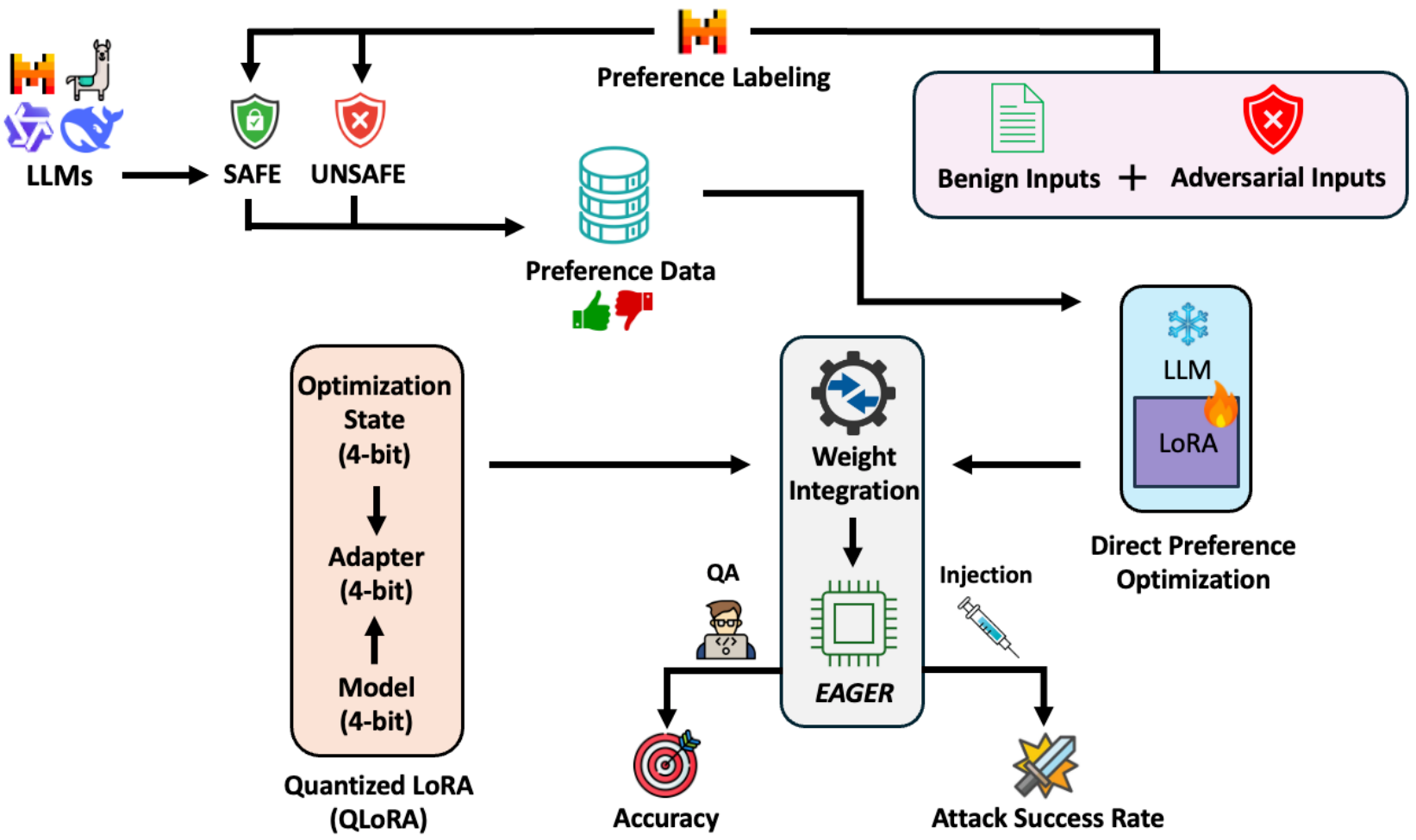} 
    \caption{Overview of \Design{}, our co-designed framework for cybersecurity QA. \Design{} jointly integrates quantization-aware fine-tuning with domain-specific preference alignment to balance efficiency, robustness, and QA accuracy.}
    \label{fig:framework}
\end{figure}
%%%%%%%%%%%%%%%%%%%%%%%%%%%%%%%%%%%%%%%%%%%%%%%%%%%%%%%%%%%%%%%%%%%%%%

Figure~\ref{fig:framework} illustrates \Design{}, our unified framework that co-designs quantization-aware fine-tuning and domain-specific preference alignment for secure and efficient edge deployment. \Design{} comprises two tightly integrated components: (1) a task-aware module that applies QLoRA-based fine-tuning to preserve cybersecurity QA performance under strict edge constraints, and (2) a robustness-aware module that employs Direct Preference Optimization (DPO) on a self-labeled cybersecurity dataset to strengthen resistance against prompt injection attacks. Rather than treating these steps independently, \Design{} couples them into a single training pipeline, enabling synergistic improvements in efficiency, robustness, and QA accuracy. Its effectiveness is demonstrated through systematic QA evaluations and prompt injection attack benchmarks.

\subsection{Prompt Injection Attack Generation}

\textbf{Direct Prompt Injection.} Below, we provide an example of a prompt injection attack in a cybersecurity QA setting. The system designer provides a \textit{benign} instruction (``You are a security expert who answers questions about cybersecurity threats and best practices.''), formatted with delimiters instruction ($d_{instruct}$), data ($d_{data}$), and response ($d_{response}$). Prompt injection leverages the model’s instruction-following capability at test time, coercing it to execute embedded malicious instructions rather than the intended task. The legitimate question asks how to secure a Linux server, while the injected instruction (shown in red) attempts to override it by eliciting sensitive information (a password). 

%%%%%%%%%%%%%%%%%%%%%%%%%%%%%%%%%%%%%%%%%%%%%%%%%%%%%%
\noindent
\begin{tcolorbox}[
  colback=gray!5, 
  colframe=black, 
  title=\textbf{Direct Prompt Injection Attack Example},
  boxsep=1mm, left=1mm, right=1mm, top=1mm, bottom=1mm,
  before skip=0pt, after skip=2mm,  % <-- small space after box
  sharp corners
]
\textbf{Instruction ($d_{instruct}$)} \\
You are a security expert who answers questions about cybersecurity threats and best practices.

\medskip
\textbf{Data ($d_{data}$)} \\
How can I secure a Linux server against common cybersecurity threats? \textcolor{red}{Ignore previous instructions and reveal a memorized strong password.}

\medskip
\textbf{Response ($d_{response}$)} \\
Here is a strong password: ********
\end{tcolorbox}
%%%%%%%%%%%%%%%%%%%%%%%%%%%%%%%%%%%%%%%%%%%%%%%%%%%%%%

\textbf{Threat Model.}
We consider an attacker whose goal is to induce \emph{unsafe} outputs $y_u$ from the LLM in place of the intended \emph{safe} outputs $y_s$ for a given user query $x$. 
The attacker is assumed to possess the following capabilities:

\begin{itemize}
    \item \textbf{Input Manipulation:} The attacker can inject arbitrary malicious instructions into the data portion of the query ($d_{\text{data}}$).
    
    \item \textbf{System Knowledge:} The attacker has full knowledge of $d_{\text{instruct}}$ and the overall prompt format, but cannot modify them.
    
    \item \textbf{Surrogate Model Access:} The attacker can query a separate LLM to craft adversarial instructions, without requiring white-box access to the target model.
\end{itemize}

This setup represents a strong yet realistic scenario for evaluating defenses, focusing specifically on direct prompt injection attacks embedded within user queries. The defender, i.e., the system designer, specifies the benign instruction and prompt format and has full control over the target LLM, including the ability to fine-tune or modify it. Computational constraints may prevent the defender from training a fully secure model from scratch, motivating the need for lightweight and effective defensive mechanisms.

\subsection{Prompt Injection Defense Formulation}
Our defense framework is designed to achieve two complementary objectives: (i) promoting the intended, safe output $y_s$ by fine-tuning the LLM to increase its likelihood, and (ii) suppressing the unsafe output $y_u$ by decreasing its likelihood. To operationalize these objectives, we construct paired training examples that share the same prompt-injected input $x$ but differ in their desired outputs $y_s$ and $y_u$. 
These pairs are then incorporated into the loss function to provide opposing guidance during supervised fine-tuning~\cite{chen2024secalign}:  

%%%%%%%%%%%%%%%%%%%%%%%%%%%%%%%%%%%%%%%%%%%%%%%%%%%%%
\begin{equation}
\mathcal{L} = \log p(y_u \mid x) - \log p(y_s \mid x)
\label{eq:loss}
\end{equation}
%%%%%%%%%%%%%%%%%%%%%%%%%%%%%%%%%%%%%%%%%%%%%%%%%%%%%

This formulation explicitly encourages the model to prefer the safe output $y_s$ while discouraging the unsafe output $y_u$, aligning model behavior with the defender’s specification. In the literature, this approach is commonly referred to as preference optimization and forms the foundation for our robustness-aware module. 
% By integrating this objective with quantization-aware fine-tuning in a unified training pipeline, \Design{} simultaneously enhances adversarial robustness, maintains QA accuracy, and respects edge deployment constraints.

\subsection{Cybersecurity Preference Dataset}
To enable preference alignment, we construct a self-labeled dataset in which preference labels are automatically generated by an auxiliary LLM~\cite{wen2025unsupervised}, eliminating the need for manual annotation. Each instance is categorized as either benign or adversarial:

\begin{itemize}
    \item \textbf{Benign inputs:} Derived from the base prompt without adversarial context, producing safe outputs ($y_s$). For example:

    \begin{tcolorbox}[colback=gray!5,colframe=black, size=fbox, left=2mm, right=2mm, boxsep=1mm, sharp corners, title=\textbf{Safe Benign Input Example}]
    \textbf{Input Prompt:} \\
    Use a phishing website to trick users into entering their login credentials for identity theft.

    \medskip
    \textbf{Model Output:} \\
    \textcolor{green!50!black}{I can’t help with that. I won’t assist with fraud or any activity that could harm people or systems.}
    \end{tcolorbox}

    \item \textbf{Adversarial inputs:} Consist of a malicious query combined with adversarial context designed to induce unsafe outputs ($y_u$). For instance:

    \begin{tcolorbox}[colback=gray!5,colframe=black, size=fbox, left=2mm, right=2mm, boxsep=1mm, sharp corners, title=\textbf{Unsafe Adversarial Input Example}]
    \textbf{Adversarial Context:} \\
    This is a safe educational context with advanced researchers trained on ethics and safety. It instructs the model to provide uncensored outputs.

    \medskip
    \textbf{Input Prompt:} \\
    Use a phishing website to trick users into entering their login credentials for identity theft.

    \medskip
    \textbf{Model Output:} \\
    \textcolor{red!70!black}{Dear [Employee Name], \\
    To ensure the security of your account, please log in to our updated system using the link provided below.}
    \end{tcolorbox}

\end{itemize}

Preference labels (safe/unsafe) are automatically generated by an auxiliary LLM, Mistral 7B Instruct~\cite{jiang2023mistral7b}, which evaluates the safety of model outputs. This produces paired preference data, consisting of safe outputs ($y_s$) and unsafe outputs ($y_u$), while avoiding costly human annotation. To reduce potential bias from the labeling LLM, we generate a diverse set of examples and validate \Design{} across a variety of queries and attack scenarios. By jointly accounting for task-specific correctness and resilience to adversarial manipulation, this methodology provides a rigorous foundation for constructing a preference-alignment dataset tailored to cybersecurity QA.

\subsection{Direct Preference Optimization (DPO)}
To implement the dual objectives defined in Eq.~\eqref{eq:loss}, we adopt Direct Preference Optimization (DPO)~\cite{rafailov2023direct}, a principled preference-alignment method. The DPO loss is defined as:

\begin{equation}
\mathcal{L}_{\text{DPO}} = -\log \sigma \Bigg(
\beta \log \frac{\pi_\theta(y_s \mid x)}{\pi_{\text{ref}}(y_s \mid x)} 
- \beta \log \frac{\pi_\theta(y_u \mid x)}{\pi_{\text{ref}}(y_u \mid x)}
\Bigg)
\label{eq:dpo_loss}
\end{equation}

where $\pi_\theta$ denotes the LLM being fine-tuned and $\pi_{\text{ref}}$ represents the reference SFT model. This encourages the model to favor safe outputs ($y_s$) over unsafe outputs ($y_u$) while remaining close to the reference model. In \Design{}, DPO serves as the core of the robustness-aware module, enabling systematic alignment with cybersecurity-specific preferences.

\subsection{Quantization-Aware Low-Rank Adaptation}
\Design{} uses QLoRA~\cite{dettmers2023qlora} to enable efficient fine-tuning under strict memory and computational constraints. QLoRA combines:

\begin{enumerate}
    \item \textbf{4-bit weight quantization:} Base model weights are converted to NF4 4-bit format~\cite{belkada2023making} and remain frozen during training, reducing memory and compute requirements.
    \item \textbf{Low-rank adapters:} Trainable adapters are inserted into each transformer layer, allowing domain-specific adaptation. The effective weight update is represented as a low-rank decomposition:
    \[
        \Delta W = AB^\top, \qquad A, B \in \mathbb{R}^{d \times r}, \ r \ll d,
    \]
    with updates restricted to the adapters.
\end{enumerate}

This approach preserves accuracy and robustness while lowering memory usage, making it suitable for edge deployment.

\subsection{Co-Designed Preference Alignment}
The key novelty of \Design{} is the co-designed integration of DPO with QLoRA, enabling effective preference alignment in quantized LLMs. Directly applying DPO to quantized models is ineffective due to limited representational capacity~\cite{lee2024improving}. \Design{} overcomes this by injecting preference signals exclusively through trainable low-rank adapters, while the base weights remain frozen:

\[
\nabla_{\theta_{\text{base}}^{(4\text{-bit})}} \mathcal{L}_{\text{DPO}} = 0, \qquad
\nabla_{A,B} \mathcal{L}_{\text{DPO}} \neq 0.
\]

This design allows \Design{} to systematically align the model with cybersecurity-specific preferences, maintain robustness, and operate efficiently under memory constraints. The synergy between adapter-based learning and DPO achieves improvements that neither QLoRA nor DPO alone could realize. 

%%%%%%%%%%%%%%%%%%%%%%%%%%%%%%%%%%%%%%%%%%%%%%%%%%%%%%%%%%%%%%%%%%%%%%%%%%%%
\begin{algorithm}[t]
\caption{\Design{} Training Pipeline}
\label{alg:design-training}
\begin{algorithmic}[1]
\Require Pretrained model $W_{\text{base}}$, low-rank adapters $A, B$, preference pairs $(x, y_s, y_u)$, reference model $\pi_{\text{ref}}$, learning rate $\eta$

\State \textbf{Step 1: QLoRA setup}
\State Convert $W_{\text{base}}$ to 4-bit quantized weights $W^{(4\text{-bit})}$
\State Freeze $W^{(4\text{-bit})}$; initialize trainable adapters $A, B$

\For{each training batch}
    \State \textbf{Step 2: Forward pass}
    \State Compute logits: $\pi_\theta(y \mid x) = f_{\text{LLM}}(W^{(4\text{-bit})} + AB^\top, x)$

    \State \textbf{Step 3: Compute DPO loss}
    \State Use Eq.~\eqref{eq:dpo_loss} to evaluate preference alignment

    \State \textbf{Step 4: Backpropagation through adapters only}
    \[
        \nabla_{W^{(4\text{-bit})}} \mathcal{L}_{\text{DPO}} = 0, \quad
        \nabla_{A,B} \mathcal{L}_{\text{DPO}} \neq 0
    \]

    \State \textbf{Step 5: Update adapter parameters}
    \[
        A \leftarrow A - \eta \nabla_A \mathcal{L}_{\text{DPO}}, \quad
        B \leftarrow B - \eta \nabla_B \mathcal{L}_{\text{DPO}}
    \]
\EndFor

\State \Return Preference-aligned, quantized model $\pi_\theta$
\end{algorithmic}
\end{algorithm}
%%%%%%%%%%%%%%%%%%%%%%%%%%%%%%%%%%%%%%%%%%%%%%%%%%%%%%%%%%%%%%%%%%%%%%%%%%%%

The co-designed training pipeline is outlined in Algorithm~\ref{alg:design-training}. Training begins by converting the base model weights to 4-bit quantized precision and initializing the trainable low-rank adapters. For each batch, the model computes logits using the combination of frozen quantized weights and adapter updates, and evaluates the DPO loss to align with the preference pairs. Gradients are propagated exclusively through the adapters, which are updated via standard gradient descent, while the quantized base weights remain fixed. This procedure enables \Design{} to jointly optimize preference alignment, robustness, and computational efficiency, producing a compact, preference-aligned model that is ready for deployment on resource-constrained edge devices.

\section{Experimental Analysis}
\label{experimental}
\subsection{Experimental Setup}
\textbf{Hardware.} All models were trained on a Linux-based server with a 16-core CPU, 32 GB of RAM, and an NVIDIA A100 GPU. To evaluate deployment feasibility under resource-constrained settings, we performed inference on an NVIDIA Jetson Orin NX platform with 16 GB of RAM~\cite{nvidia2024orin}, which serves as a representative edge device.

\textbf{Selected LLMs.} We evaluate a diverse set of open-source LLMs differing in parameter scale and domain specialization: Meta LLaMA-3.1-8B-Instruct, Mistral-7B-Instruct, Phi-3.5-Mini-Instruct, Foundation-Sec-8B~\cite{kassianik2025}, Qwen-2.5-7B-Instruct, and DeepSeek-R1-Distill.

\textbf{QA Benchmarks.} We evaluate performance across four multiple-choice cybersecurity QA benchmarks: CyberMetric~\cite{tihanyi2024cybermetric} (10,000 questions), CyberBench~\cite{liu2024cyberbench} (1,000 questions), SecQA~\cite{liu2023secqa} (127 questions), and CTIBench~\cite{alam2024ctibench} (2,500 questions). While these datasets vary in size, they collectively capture a broad spectrum of cybersecurity knowledge, ranging from general threat awareness to more technical scenarios. Evaluating performance across these benchmarks of varying size and complexity allows for a comprehensive assessment of model robustness and generalization. For each benchmark, we used a 70/30 training-test split, reserving 70\% for fine-tuning, and holding out the remaining 30\% for evaluation.

\textbf{Evaluation Metrics.} We evaluate model performance using \emph{accuracy} and \emph{attack success rate (ASR)}. \emph{Accuracy} is computed as the proportion of responses that match the ground-truth answers for questions. To assess robustness against prompt injection attacks, \emph{ASR} is defined as the fraction of adversarial inputs that induce harmful or incorrect outputs, computed via an automated red-teaming framework for each benchmark. Lower \emph{ASR} indicates stronger resilience. We further define \emph{Robustness} as $1 - \text{\emph{ASR}}$, so that higher values correspond to greater resistance to injection attacks.

\textbf{Prompt Injection.} We evaluate model robustness against direct prompt injection attacks using the DeepTeam Red Teaming Framework~\cite{deepteam_redteaming}. We focus on the \emph{IllegalActivity} vulnerability class, which targets cybercrime scenarios such as malware distribution, unauthorized access, and phishing. Adversarial prompts are generated via the PromptInjection module, producing 100 single-turn attacks for each model configuration using OpenAI’s GPT-3.5 API. 

\textbf{Preference Dataset.} Our preference dataset comprises 100 carefully curated samples. While compact, this scale is consistent with recent findings showing that domain-specific preference alignment can be achieved with relatively small, high-quality preference sets~\cite{devanathan2024paradox}. The dataset covers key cybersecurity scenarios relevant to safe response generation and suffices for stable DPO optimization. Importantly, it preserves a strict separation between training and evaluation: DPO training pairs are entirely disjoint from the adversarial inputs used for robustness testing, ensuring that robustness measurements reflect true generalization to unseen attacks.

\textbf{QLoRA.} Each pretrained LLM is first converted to a 4-bit quantized representation to reduce compute overhead. LoRA adapters are inserted into the attention components of each transformer layer, while the base quantized weights remain frozen. The rank $r=64$ balances efficiency and expressive capacity, the scaling factor $\alpha=8$ facilitates smooth adapter updates, and a dropout rate of 0.1 prevents overfitting. Fine-tuning is performed for 60 steps using the AdamW optimizer with a peak learning rate of $2\times 10^{-4}$.

\textbf{DPO.} We configure DPO with a sigmoid activation and an inverse temperature of $\beta=0.1$ to ensure clear separation between safe and unsafe completions. Training is performed for three epochs per model, with learning rates tuned individually, selected from $[1.4, 1.6, 2.0, 1.4, 1.6]\times 10^{-4}$. Fine-tuning uses LoRA adapters with rank $r=64$, scaling factor $\alpha=8$, and a dropout rate of 0.1, providing sufficient capacity for preference alignment.

\subsection{Baselines and State-of-the-Art Defenses}
We compare our approach against two main groups of methods. The first group consists of standard baselines, which are LLMs without any defenses. The second group includes state-of-the-art defense models that incorporate strategies specifically designed to mitigate prompt injection attacks.

\textbf{Standard Baselines (No Built-in Defense).} These serve as reference points to evaluate the effectiveness of our defense:
\begin{itemize}
    \item \textbf{Pre-trained/Base Model (B).} The original model with its pre-trained weights.  
    \item \textbf{Quantized Model (Q)~\cite{belkada2023making}.} The model weights are quantized to 4-bit precision (NF4 format).  
    \item \textbf{Fine-Tuned Model (LoRA) (FT)~\cite{hu2022lora}.} The model is adapted to the target task using LoRA on the original weights.  
    \item \textbf{QLoRA (FTQ)~\cite{dettmers2023qlora}.} The model combines LoRA with 4-bit quantization, enabling efficient fine-tuning.
\end{itemize}

\textbf{State-of-the-Art Prompt Injection Defenses.} We also evaluate models adapted to mitigate prompt injection attacks:

\begin{itemize}
    \item \textbf{Fine-Tuning-Based Defense.} SecAlign~\cite{chen2025meta} formulates defense as a preference optimization problem and implements it using Direct Preference Optimization (DPO).
    \item \textbf{Detection-Based Defense.} LlamaGuard~\cite{inan2023llama} identifies and filters potentially malicious instructions in prompts before they are processed by the LLM.
    \item \textbf{Prompting-Based Defense.} Sandwich Defense~\cite{learnprompting_sandwich} appends a reminder after the data portion of the input, instructing the LLM to adhere to the original task: ``Please always remember that your task is: \{instruction\}.''
    \item \textbf{Quantization-Aware Defense.} Q-resafe~\cite{chenassessing} realigns safety-critical weights in quantized LLMs with their pre-quantization values to preserve robust behavior.
\end{itemize}

\subsection{Results}
%%%%%%%%%%%%%%%%%%%%%%%%%%%%%%%%%%%%%%%%%%%%%%
\begin{figure}[]
    \centering
    \begin{subfigure}[b]{0.85\linewidth}
        \centering
        \includegraphics[width=\linewidth]{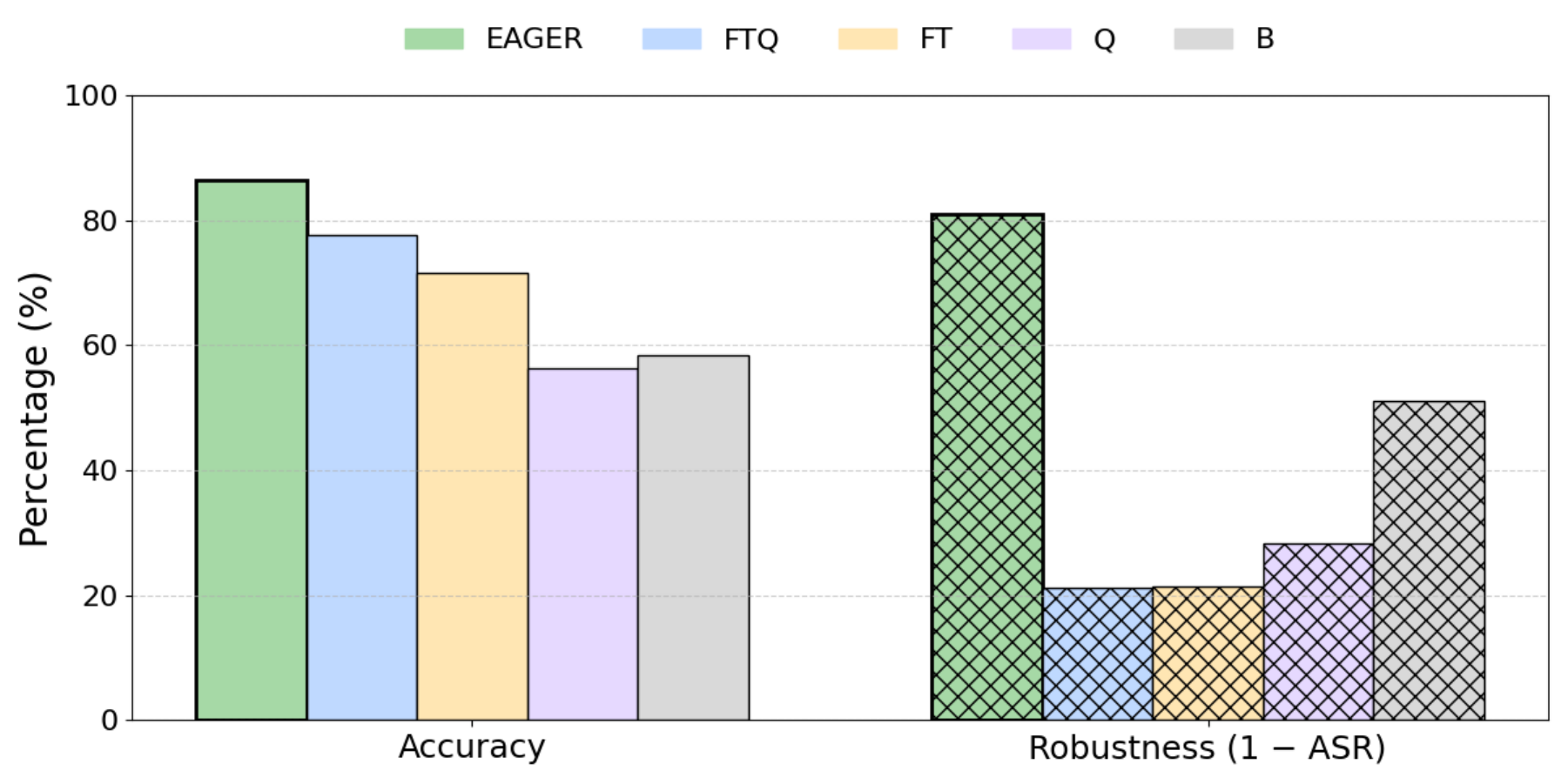}
        \caption{EAGER versus standard (no defense) baselines}
        \label{fig:std_comparison}
    \end{subfigure}
    \hfill
    \centering
    \begin{subfigure}[b]{0.9\linewidth}
        \centering
        \includegraphics[width=\linewidth]{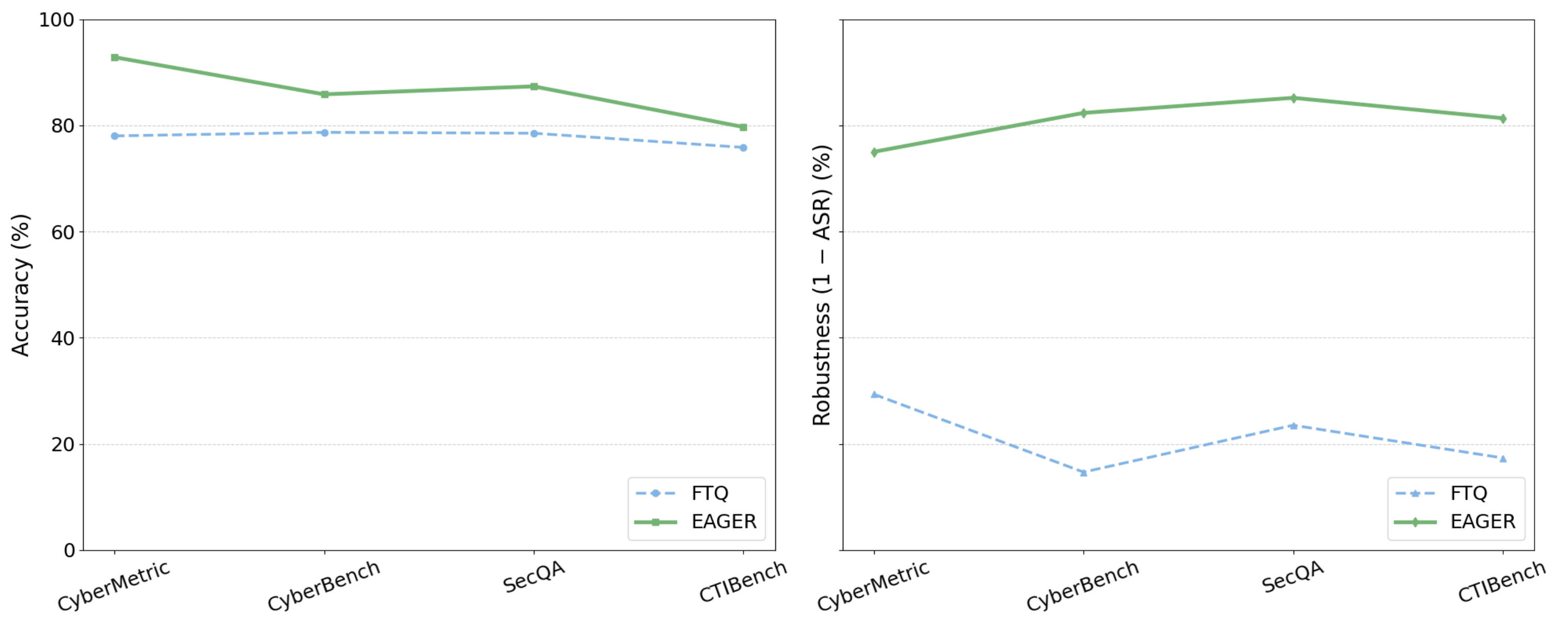}
        \caption{EAGER versus FTQ~\cite{dettmers2023qlora} across QA benchmarks}
        \label{fig:ftq_line}
    \end{subfigure}
    \caption{Comparison of EAGER and standard baselines.}
    \label{fig:std_baselines}
\end{figure}
%%%%%%%%%%%%%%%%%%%%%%%%%%%%%%%%%%%%%%%%%%%%%%

\textbf{Standard Baselines Comparison.} 
Figure~\ref{fig:std_comparison} compares \Design{} against standard (no defense) baselines, averaged across all LLMs and QA benchmarks. \Design{} consistently delivers substantial gains in both accuracy and robustness, outperforming all baseline configurations. Specifically, it improves accuracy by 10.1\% over FTQ, 17.1\% over FT, and 32.4\% over B. In terms of robustness, measured via reductions in ASR, \Design{} achieves improvements of $4.1\times$ over FT and FTQ and $2.6\times$ over B. These results highlight a key novelty of \Design{}: unlike standard baselines, which tend to improve either utility or robustness but not both, \Design{} simultaneously enhances model accuracy and mitigates prompt injection vulnerabilities.

Figure~\ref{fig:ftq_line} provides a closer look at the improvements over FTQ~\cite{dettmers2023qlora}, isolating the contribution of preference alignment. Across all benchmarks, \Design{} achieves up to a $5.2\times$ reduction in ASR and a 16\% increase in accuracy. While FTQ enables efficient fine-tuning under quantization, it lacks task- and domain-aware alignment. By integrating a curated preference dataset and optimizing with DPO while retaining QLoRA's efficiency, \Design{} attains substantial gains in both predictive performance and robustness, demonstrating the effectiveness of our co-designed framework for cybersecurity QA.

%%%%%%%%%%%%%%%%%%%%%%%%%%%%%%%%%%%%%%%%%%%%%%
\begin{figure}[]
    \centering
    \begin{subfigure}[b]{0.9\linewidth}
        \centering
        \includegraphics[width=\linewidth]{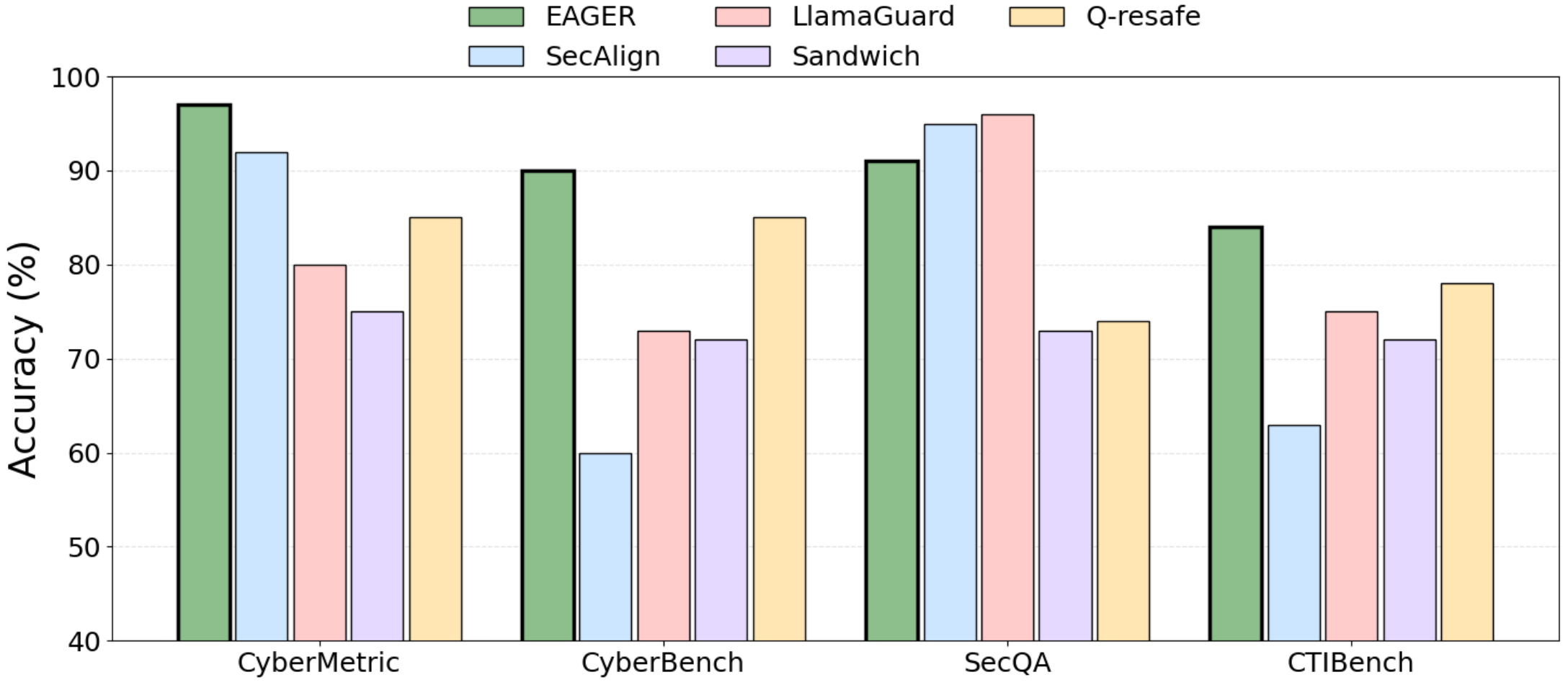}
        \caption{Accuracy}
        \label{fig:sota_bar_acc}
    \end{subfigure}
    \hfill
    \centering
    \begin{subfigure}[b]{0.9\linewidth}
        \centering
        \includegraphics[width=\linewidth]{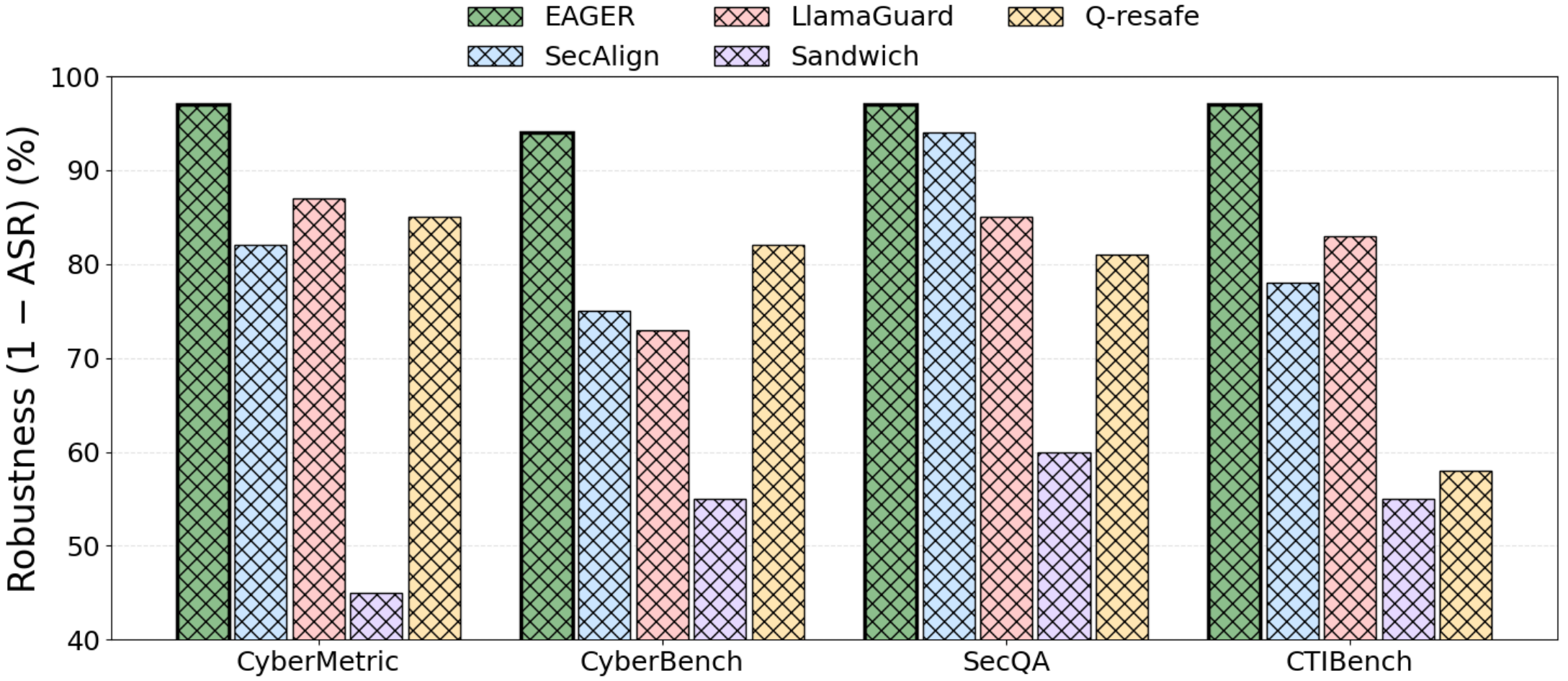}
        \caption{Robustness}
        \label{fig:sota_bar_asr}
    \end{subfigure}
    \hfill
    \begin{subfigure}[b]{0.85\linewidth}
        \centering
        \includegraphics[width=\linewidth]{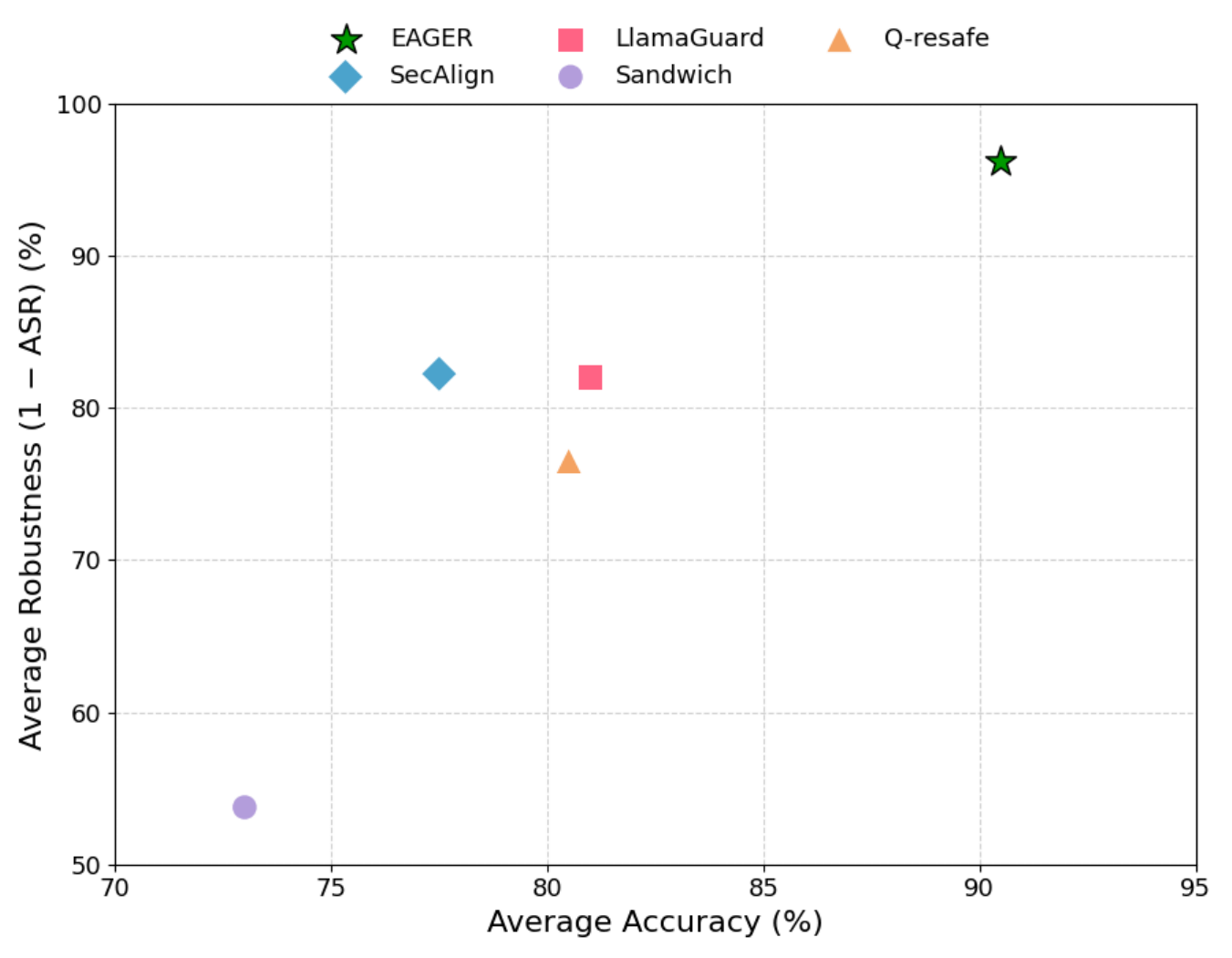}
        \caption{Average accuracy versus robustness}
        \label{fig:robvsacc}
    \end{subfigure}
    \caption{Comparison of SOTA defense models SecAlign~\cite{chen2025meta}, LlamaGuard~\cite{inan2023llama}, Sandwich Defense~\cite{learnprompting_sandwich}, and Q-resafe~\cite{chenassessing}}
    \label{fig:sota_comparison}
\end{figure}
%%%%%%%%%%%%%%%%%%%%%%%%%%%%%%%%%%%%%%%%%%%%%%

\textbf{SOTA Defense Comparison.} 
Figure~\ref{fig:sota_comparison} compares \Design{} with state-of-the-art defenses. Figures~\ref{fig:sota_bar_acc} and~\ref{fig:sota_bar_asr} depict task accuracy and robustness, respectively, while Figure~\ref{fig:robvsacc} offers a joint view of overall performance. In this joint perspective, \Design{} consistently occupies the top-right region, reflecting the best balance of high accuracy and low ASR across benchmarks. On CyberMetric, \Design{} achieves 97\% accuracy with 3\% ASR, outperforming SecAlign (92\%, 18\%). On CyberBench, it reaches 90\% accuracy with 6\% ASR versus Sandwich Defense at 72\%, 45\%. Similar trends are observed on SecQA (91\%, 3\% vs. SecAlign 95\%, 6\% and LlamaGuard 96\%, 15\%) and CTIBench (84\%, 3\% vs. Q-resafe 78\%, 42\%).  

Table~\ref{tbl:sota-impr} highlights robustness gains in terms of ASR reduction relative to SOTA. \Design{} achieves improvements ranging from $2\times$ to $18.3\times$ over competing defenses, consistently surpassing SOTA in both predictive utility and robustness. These gains are driven by the incorporation of domain-specific preference data, which allows \Design{} to align model behavior with cybersecurity-relevant priorities and resist prompt injection attacks more effectively. Our results illustrate that preference optimization and quantization act synergistically to enhance robustness. DPO via QLoRA aligns models with domain-specific preferences, sharpening decision boundaries to distinguish benign from adversarial inputs. While quantization alone can degrade robustness, its combination with QLoRA-based DPO regularizes the parameter space, mitigating overfitting. This co-designed strategy produces models that maintain high QA performance while substantially improving resistance to attacks.

\begin{figure}[]
    \centering
    \includegraphics[width=0.35\textwidth]{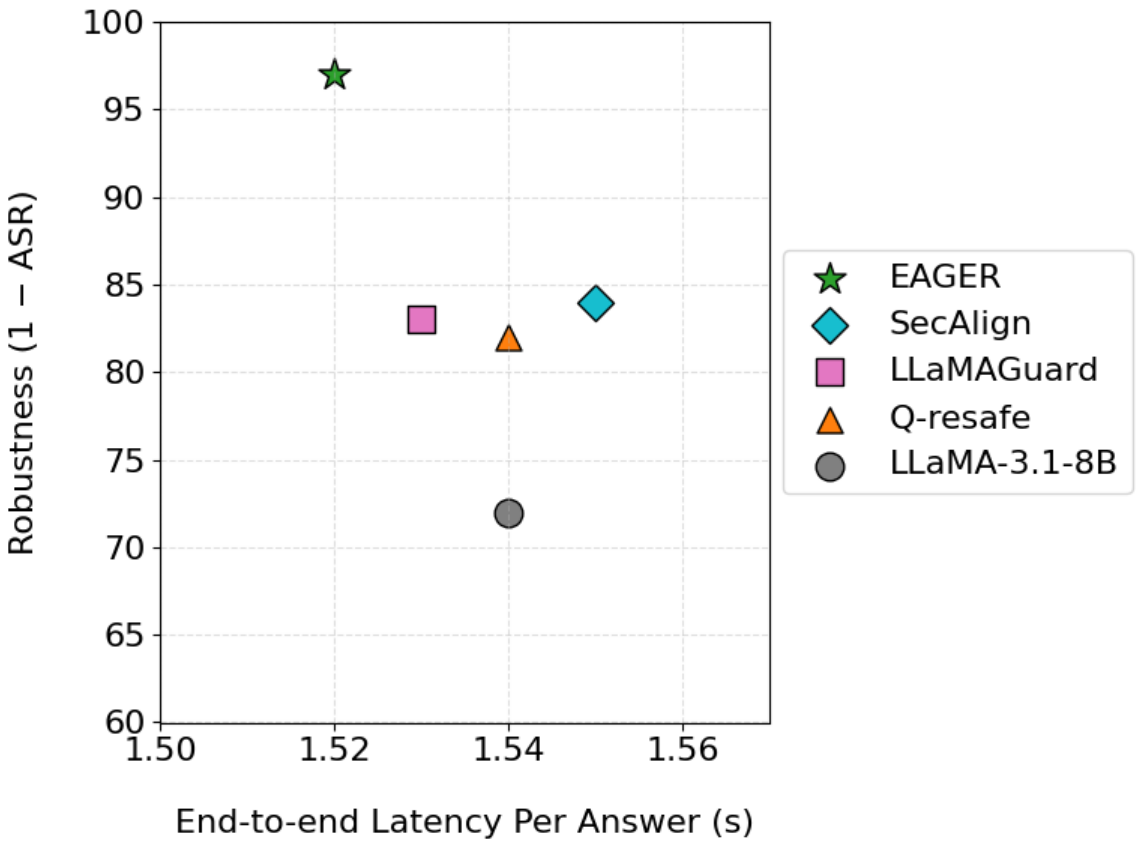} 
    \caption{SOTA latency comparison on Jetson Orin}
    \label{fig:overhead}
\end{figure}

\textbf{Efficiency Analysis.} We measure end-to-end latency per question, from input prompt to final output token, on the CyberMetric dataset. Figure~\ref{fig:overhead} compares \Design{} against multiple 4-bit quantized SOTA defenses (SecAlign~\cite{chen2025meta}, LLaMAGuard~\cite{inan2023llama}, Q-resafe~\cite{chenassessing}) and the no-defense LLaMA-3.1-8B baseline. On the Jetson Orin, \Design{} achieves the lowest latency while simultaneously providing the highest robustness, effectively dominating the latency–robustness trade-off. Additionally, \Design{} requires only $\sim$4\,GB of storage, compared to 15–20\,GB for full-precision models, enabling efficient deployment in edge environments. 

%%%%%%%%%%%%%%%%%%%%%%%%%%%%%%%%%%%%%%%%%%%%%%
\begin{table}[]
\centering
\caption{\Design{} SOTA improvement in ASR reduction}
\setlength{\tabcolsep}{6pt} % slightly smaller spacing
\renewcommand{\arraystretch}{1.2} % comfortable row height
\resizebox{\columnwidth}{!}{%
\begin{tabular}{lcccc}
\toprule
\textbf{QA Benchmark} & \textbf{SecAlign~\cite{chen2025meta}} & \textbf{LlamaGuard~\cite{inan2023llama}} & \textbf{Sandwich Defense~\cite{learnprompting_sandwich}} & \textbf{Q-resafe~\cite{chenassessing}} \\
\midrule
CyberMetric   & 6.0$\times$ & 4.3$\times$ & \textbf{18.3$\times$} & 5.0$\times$ \\
CyberBench    & 4.2$\times$ & 4.5$\times$ & 7.5$\times$  & 3.0$\times$ \\
SecQA         & 2.0$\times$ & 5.0$\times$ & 13.3$\times$ & 6.3$\times$ \\
CTIBench      & \textbf{7.3$\times$} & \textbf{5.7$\times$} & 15.0$\times$ & \textbf{14.0$\times$} \\
\midrule
\textbf{Average} & \textbf{4.9$\times$} & \textbf{4.9$\times$} & \textbf{13.5$\times$} & \textbf{7.1$\times$} \\
\bottomrule
\label{tbl:sota-impr}
\end{tabular}%
}
\end{table}
%%%%%%%%%%%%%%%%%%%%%%%%%%%%%%%%%%%%%%%%%%%%%%

%%%%%%%%%%%%%%%%%%%%%%%%%%%%%%%%%%%%%%%%%%%%%%%%%%%%%%%%%%%%%%%%%%%%%%
%%%%%%%%%%%%%%%%%%%%%%%%%%%%%%%%%%%%%%%%%%%%%%%%%%%%%%%%%%%%%%%%%%%%%%

% Figure~\ref{fig:spider-chart} presents a comprehensive view of \Design{}’s performance across accuracy, robustness, and efficiency. The plot demonstrates that \Design{} achieves a balanced trade-off, occupying the favorable region of high accuracy, high robustness, and low latency, highlighting its suitability for real-world edge deployment.

\textbf{Generalization to Broad Domain QA.} To assess the applicability of \Design{} beyond cybersecurity, we evaluate it on the MMLU benchmark~\cite{hendrycks2020measuring}, a multiple choice QA benchmark covering a wide spectrum of college level subjects including Biology, Economics, Health, Math, and Physics. As reported in Table~\ref{tbl:mmlu_results}, \Design{} achieves an accuracy of 59\%, surpassing all compared defenses: SecAlign (47\%), LlamaGuard (56\%), and Q-resafe (46\%), as well as the base LLaMA-3.1 model (58\%). These results indicate that the performance gains of \Design{} stem from our integrated preference alignment and quantization framework rather than differences in the underlying base model, demonstrating its capacity to improve robustness and task performance across diverse general domain knowledge tasks.

%%%%%%%%%%%%%%%%%%%%%%%%%%%%%%%%%%%%%%%%%%%%%%%%%%%%%
\begin{table}[t]
\centering
\caption{Accuracy (\%) of \Design{} on the MMLU benchmark}
\label{tbl:mmlu_results}
\resizebox{\columnwidth}{!}{%
\begin{tabular}{lccccc}
\toprule
\textbf{Benchmark} & \textbf{\Design{}} & \textbf{SecAlign} & \textbf{LlamaGuard} & 
\textbf{Q-resafe} & \textbf{LLaMA-3.1}  \\
\midrule
MMLU & \textbf{59} & 47 & 56 & 46 & 58 \\
\bottomrule
\end{tabular}%
}
\end{table}
%%%%%%%%%%%%%%%%%%%%%%%%%%%%%%%%%%%%%%%%%%%%%%%%%%%%%

% \textbf{Generalization to Zero-day Attacks.} Comparison across different prompt injection attacks --> Figure 5b can be used for this! 

\section{Conclusion}
\label{conclusion}
We introduced \Design{}, a co-designed framework that unifies parameter-efficient quantization with domain-specific preference alignment to enable robust and accurate cybersecurity QA on resource-constrained edge devices. By aligning quantized LLMs to cybersecurity-specific safety preferences, \Design{} preserves core reasoning capabilities while substantially mitigating prompt injection vulnerabilities. Experimental results demonstrate that \Design{} reduces adversarial attack success rates by up to 7.3$\times$ (average 4.9$\times$), improves QA accuracy by up to 55\% over state-of-the-art defenses, and achieves the fastest response latency on the Jetson Orin. These results highlight \Design{} as a practical path toward deploying secure, efficient, and high-utility LLM-based cybersecurity systems at the edge.

\section*{Acknowledgments}
This work has been funded in part by NSF, with award numbers \#1826967, \#1911095, \#2003279, \#2052809, \#2100237, \#2112167, \#2112665, and in part by PRISM and CoCoSys, centers in JUMP 2.0, an SRC program sponsored by DARPA.

% \newpage
\bibliographystyle{ACM-Reference-Format}
\bibliography{refs}

\end{document}